\documentclass[pt10,fleqn,usenatbib]{mnras}
\usepackage{newtxtext,newtxmath}
\usepackage[T1]{fontenc}
\usepackage{ae,aecompl}
\usepackage{comment}
\usepackage{graphicx}	
\usepackage{amsmath}	
\usepackage{amssymb}	
\title[Ceres' crust convection]{Thermal convection in the crust of the dwarf planet (1) Ceres}

\author[M. Formisano et al.]{
M. Formisano,$^{1}$\thanks{E-mail: michelangelo.formisano@inaf.it (MF)}
C. Federico,$^{1}$
J. Castillo-Rogez,$^{2}$
M.C. De Sanctis$^{1}$
and G. Magni$^{1}$
\\
$^{1}$INAF-IAPS, Via del Fosso del Cavaliere 100, 00133, Rome, Italy\\
$^{2}$Jet Propulsion Laboratory, California Institute of Tecnhology, Pasadena, CA, USA\\
}

\date{Accepted XXX. Received YYY; in original form ZZZ}

\begin{document}
\label{firstpage}
\pagerange{\pageref{firstpage}--\pageref{lastpage}}
\maketitle

\begin{abstract}
Ceres is the largest body in the Main Belt, and it is characterized by a large abundance of water ice in its interior. This feature is suggested by its relatively low bulk density (2162 kg m$^{-3}$), while its partial differentiation into a rocky core and icy crust is suggested by several geological and geochemical features: minerals and salts produced by aqueous alteration, icy patches on the surface, lobate morphology interpreted as surface flows.
In this work we explore how the composition can influence the characteristics of thermal convection in the crust of Ceres. Our results suggest that the onset of thermal convection is difficult and when it occurs it is short lived and this could imply that Ceres preserved deep liquid until present, as recent suggested by the work of Castillo-Rogez et al.. Moreover, cryovolcanism could be driven by diapirism (chemical convection) rather than thermal convection. 
\end{abstract}




\section{Introduction}
The in-depth study of the surface of  the dwarf planet Ceres based on the observations returned by the Dawn spacecraft over the past few years, has highlighted the extensive presence of minerals produced by aqueous alteration (e.g. VIR observations) \citep{DeSanctis16,Combe16,Ciarniello17,Raponi2018a,Carrozzo2018,McSween18,Combe18}, geomorphological evidence for ground ice \citep{Schmidt}, and an ice table via hydrogen detection \citep{Prettyman2017} all indicating that a large quantity of water is present both on the surface and in the interior of this planetary object. Moreover, past studies, e.g. \citet{MCCord05,Castillo2010,McCord2011} investigated the role of this huge quantity of water in the thermodynamic evolution of Ceres.

The inversion of the gravity and topography data returned by Dawn \citep{Park16,Fu2017,Ermakov17} suggests a two-layer structure for Ceres' interior: an inner rocky layer with a density of $\simeq$2400 kg m$^{-3}$ and a crust (40 km thick) with a density of $\simeq$1300 kg m$^{-3}$. Crustal density estimates vary among studies depending on underlying assumptions: \citet{Konopliv18} derived the range 1200-1600 kg m$^{-3}$, whereas \citet{King18} estimated a crustal density lower than 1300 kg m$^{-3}$, consistent with \citet{Ermakov17}.  All these estimates indicate a partially differentiated structure with abundant water ice in the crust. Based on mechanical strength observations, \citet{Bland2016} imposed further constraints on crustal composition, suggesting no more than 40 vol$\%$ of a weak phase (water ice and/or porosity).

\citet{Formisano16b} suggested that, in the case of a pure water ice mantle, the primordial crust would not survive until present (only until 3 Ga after formation), since it experienced gravitational overturn. On the other hand, a 1-km thick crust with a low volume fraction of ice above a muddy mantle could be characterized by a stability time span compatible with the lifetime of the Solar System. However, even if the muddy mantle scenario could be still considered a possibility, a 1-km thick crust seems to be unlikely given the recent Dawn observations.

The abundant water ice in Ceres raises the possibility that cryovolcanism could occur in Ceres' crust. This paper investigates this question and tests if that process could be triggered by thermal convection. 
Cryovolcanism has been suggested as responsible for the formation of domes across Ceres' surface \citep{Sori18} and in particular for Ahuna Mons, an isolated, 4-km high mountain located at 10$^\circ$S. This outstanding feature has been interpreted as a cryovolcanic extrusive edifice \citep{Ruesch16}. Cryovolcanism could be linked to thermal convection that is a function of material viscosity, and thus of temperature. The viscosity can change by several orders of magnitude during the thermal/geophysical evolution of a planetary body and this affects the longevity and the shape of the convective cells. Here, we will investigate, through a 2-D finite element method (FEM) numerical modelling, how the composition of Ceres' crust can affect the onset of thermal convection, selecting two different composition on the basis of the constraints of the mean crustal density. The aim of this work is to evaluate the feasibility of the thermal convection in the current crust (or potentially also in the past - with the same temperature boundary conditions) given the temperature at the bottom set by geochemical considerations related to the possible presence of liquid water near the top of the mantle \cite{Castillo19}. Moreover, we will discuss the possibility to produce features such as Ahuna Mons with the stress induced by thermal convection. The lack of numerical modelling of thermal convection in Ceres' crust, as well as the importance of the physical consequences of this mechanism in a dwarf planet as Ceres are the main motivations behind this work. Moreover, the prospect for thermal convection, or absence thereof, in Ceres bears important implications for the until present preservation of liquid in the dwarf planet \citep{Castillo19}, as also proposed by a recent thermophysical model \citep{Neumann20}.

The paper is structured as follow: in Section II we present the governing equations used in this paper, with the initial configurations adopted and the viscosity calculation; in Section III we report simulation results, and finally, in Section IV, we report the conclusions. An Appendix with additional information is provided.

\section{2-D Mathematical Model}
\subsection{Governing Equations}
We applied a 2-D finite element method (FEM) model in order to solve the Navier-Stokes equations coupled with the heating equation, by using the software COMSOL Multiphysics. We solve the following system of equations:
\begin{align}\label{eq:NS}
\rho\left[\frac{\partial \vec u}{\partial t} + \vec u \cdot \vec \nabla u\right] &= \vec \nabla \cdot \left(-p\vec I + \vec W\right) + \rho \vec g\\
\rho\left(\vec \nabla \cdot \vec u\right) &= 0\\ 
\vec W &= \eta(T)\left[\vec \nabla u + \left(\vec \nabla u\right)^T\right],
\end{align} 
where $t$ is the time, $\rho$ is the density, $\vec u$ the convective velocity, $\eta(T)$ the dynamic viscosity, $T$ the temperature, $p$ the pressure, $g$ the gravity acceleration and $\vec W$ is the stress.

Some approximations are used: I) we consider the density as a function of temperature only in the buoyancy term (Boussinesq approximation); II) incompressible fluid (i.e. $\rho \left(\nabla \cdot u\right)=0$); III) no sink or source of new materials are included. However, we can neglect the momentum term due to the large value of the dynamic viscosity which reflects in a near infinity Prandtl number (see Eq.(A8) in the Appendix).

Navier-Stokes equations are coupled with the heating equation:

\begin{equation}\label{eq:heat}
\rho c_p \left[\frac{\partial T}{\partial t} + \vec u \cdot \vec \nabla T\right] + K\vec \nabla\cdot\vec\nabla T = 0,
\end{equation}
where $c_p$ is the specific heat and K the thermal conductivity. 
We model the crust  as a mixture of water ice, clathrate hydrate, rock (antigorite) and salt (hydrohalite) as in \citet{Castillo19}, studying the thermal evolution as a function of the volumetric percentages of such materials and temperature gradient across the shell. The thermophysical parameters adopted in this study are reported in the Appendix. Our simulations solve the dimensionless version of the above equations. See the Appendix for the non-dimensionalization procedure.

\subsection{Initial Configurations /Assumptions and Boundary Conditions}
A 2-D Cartesian geometry is adopted (see the Appendix), taking advantage of the results obtained by \citet{Vangelov94}, who showed that plane layer numerical models can be scaled to spherical geometry with no substantial errors. Moreover, the assumption of a 2-D Cartesian geometry is valid since the thickness of the crust is small compared to the radius of the body. We investigate two different configurations characterized by a particular composition, whose components are chosen according to the previously discussed observation-based studies. In Tab.\ref{tab:tab1} we report the volumetric percentage of the materials we assume in our simulations, the mean density, the mean thermal conductivity, the mean dynamic viscosity and finally the initial Rayleigh number. 
As discussed in \citet{Castillo19}, because the porosity is not well constrained, we neglect its contribution in these simulations. The simulations start with an initial temperature profile consisting in a perturbation of the linear profile, between the temperature at the base of the crust (275 K) and at the surface (170 K), expressed as a single Fourier mode (see, e.g. \cite{Solomatov07}). The initial temperature chosen at the base of the crust is suggested by the ocean composition evolution model from \citet{Castillo2018,Castillo19}, while the surface temperature corresponds to the radiation equilibrium temperature of Ceres (at equator) at its average distance from the Sun. The value of 275 K can be considered as an extreme upper bound, since in \citet{Castillo19} a temperature of 250 K is compatible with a narrow set of evolutionary scenarios.
While the surface temperature remains fixed during the simulation,
we imposed an heat flux at the bottom of the crust, whose estimation is given by \citet{Castillo19}. This heat flux decreases from an initial value of about 16 mW m$^{-2}$ (at the formation time - see Fig.1 of \citet{Castillo19}) to a current value of about 3 mW m$^{-2}$. Thus, the bottom temperature is fixed initially and changes during the time evolution.
Free slip conditions on the sides of our integration domain are imposed: this condition requires that the normal component of fluid velocity tends to zero at the wall-fluid interface while the tangential component is unrestricted.
\subsection{Viscosity}
The viscosity is the crucial parameter of this kind of simulations, since it affects the Rayleigh number and consequently the onset or not of the thermal convection as well as its evolution. There are several approaches to calculate the average viscosity of multi-phase aggregates: volume fraction weighted arithmetic mean, harmonic mean, geometric mean, logarithmic mean. A good review regarding the way to calculate the bulk viscosity is reported in \citet{Neumann20}.

The general law adopted is a temperature-dependent viscosity, largely used in literature (e.g. \cite{Thomas1987,Sterenborg13,Shoji,Rubin2014,Formisano16b}):

\begin{equation}
\eta = \eta_{0}\exp\left[A\left(\frac{T_{melt}}{T} -1\right)\right],
\end{equation}
where $\eta_{0}$ is a reference viscosity at zero pressure melting point, $A$ is a constant which depends on the activation energy and $T_{melt}$ is the melting temperature. This law is similar to that used in \citet{Solomatov07}, in which the "pre-exponential factor" is represented by $b\tau^{1-n}$, with $b$ constant linked to the original Arrhenius viscosity (e.g. \cite{Solomatov2000}) and $\tau$ the second invariant of the stress tensor.

We start the discussion of the M2 case (see Tab.\ref{tab:tab1}), whose composition is dominated by clathrates and ices being salt and rock negligible so it is not a bad approximation to consider M2 as a two components mixture. Following \cite{Grindrod2008}, the viscosity of this mixture can be calculated as:
\begin{equation}\label{eq:Grindrod}
    \eta_{mix} = x_{cla}\eta_{cla} + x_{ice}\eta_{ice},
\end{equation}
where $\eta_{cla}$ is the clathrate viscosity, $\eta_{ice}$ the ice viscosity, $x_{cla}$ the volumetric percentage of clathrate and $x_{ice}$ the volumetric percentage of ice.
However, if we calculate the viscosity of the mixture as a log average as in \citet{Formisano16b}:
\begin{equation}
    \eta_{mix} = \eta_{cla}^{x_{cla}}\eta_{ice}^{x_{ice}},
\end{equation}
the result will be not very different from that obtained with the Eq.\ref{eq:Grindrod}.

The viscosity of clathrate is considered 20 times that of water ice \citep{Durham2003}, as done in \citet{Grindrod2008}. In practice, the viscosity contrast between clathrate and ice is a function of the host species and varies with temperature \citep{Durham2003,Durham2010}. At a temperature of 250 K, the viscosity of methane clathrate may be two orders of magnitude greater than ice, whereas, the viscosity of carbon dioxide clathrate is only about 5 times greater. Since mixed clathrates are expected to form in Ceres \citep{Castillo2018}, the contrast of 20 assumed here is a reasonable assumption.
The activation energy for ice used in our work is 56 kJ/mol, which corresponds to a constant A = 25 \citep{Thomas1987,Friedson1983} and similar to the activation energy (60 kJ/mol) used in \citet{Grindrod2008}.
For clathrate, the activation energy adopted in our work is 90 kJ/mol \citep{Durham2003} and the melting temperature is considered 10 K above the ice melting temperature \citep{Burruss1981}.
As discussed in \citet{Goldsby2001} the viscosity of ice depends strongly on the grain size and stress level assumed. \citet{Rubin2014}, for example, used the following relationship for the ice viscosity:

\begin{equation}
    \eta(T) \simeq 3\times 10^{14} \left(\frac{d}{1 mm}\right)^2 \frac{T}{T_m}\exp\left[26.2\frac{T_m - T}{T}\right],
\end{equation}
where $d$ is the grain size. For grain size of the order of 1 mm, this equation is similar to that found by \citet{Thomas1987}:

\begin{equation}
    \eta(T) = \eta_0\exp\left[25\left(\frac{273}{T}-1\right)\right],
\end{equation}
with $\eta_0 = 10^{14}$ Pa s, while a grain size of the order of 0.1 mm corresponds to a reference viscosity of 10$^{12}$ Pa s.
In order to explore the dependence on the grain size and stress levels, we vary the reference viscosity ($\eta_0$) considering three cases: $10^{12}$ Pa s, $10^{13}$ Pa s and $10^{14}$ Pa s, similarly in \citet{Grindrod2008}.

In M1 case, we can approximate the four components mixture as a two components mixture: weak phase (ice) and strong phase (clathrate, salt and rock) and use Eq.\ref{eq:mumix} to calculate the viscosity of the mixture. In M1 case, strong phase represents the 70$\%$ of the mixture.

The viscosity of a mixture made only of ice and rock can be calculated by modifying the viscosity of ice through a weight-function of the volumetric percentage of rock as done in \citet{Nagel2004,Shoji}:

\begin{equation}\label{eq:mumix}
 \eta_{mix} = \frac{\eta_{ice}}{f(x_{rock})},   
\end{equation}
with $f(x_{rock}) = \left(1-\frac{x_{rock}}{\phi_{CPL}}\right)^{\beta}$, where $x_{rock}$ is the volumetric percentage of rock, $\phi_{CPL}$ is the close packing limit set at 0.74 \citep{Torquato2000} and $\beta$ ranges from 1.5 to 2.5 \citep{Nagel2004} and can be fixed at 2.0 which is the median value of that range \citep{Shoji}. The Eq.\ref{eq:mumix} is not valid if $x_{rock}$ is near $\phi_{CPL}$, but the error is not large as pointed by \citet{Rudman1992}.

This procedure leads to  results very similar to those obtained using the approach of \citet{Friedson1983} and \citet{Freeman2006}, i.e. to mulitply the ice viscosity for a function ("relative viscosity") which depends on the rock volume percentage.

We would like to note that if we calculate the bulk viscosity in M1 case using the volume fraction weighted through an  arithmetic mean, assuming $\eta_{salt}=10^{17}$ Pa s \citep{VanKeken93} and $\eta_{rock}=10^{19}$ Pa s \citep{Freeman2006} we would obtain a viscosity very close to the rock viscosity: in this case the thermal convection will be not possible.

We carried out a simulation with $\beta = 1.5$ and $\mu_{ice} = 10^{12}$ Pa s.

\section{Results}
In this section we report the results obtained by our numerical simulations; in particular, we show the temperature maps covering the whole thermal convection evolution and the 1-D temperature profile vs time at the center of the convective shell. 

\begin{table*}
	\centering
	\caption{We report the model id, the volume percentage of ice, clathrate hydrate, salt (hydrohalite), "rock" (antigorite), the mean density, the thermal conductivity and the initial dynamic viscosity of the composition and finally the initial Rayleigh number. The composition M1 is also discussed in \citet{Castillo19}, while composition M2 could be seen as an "extreme" case, being slightly outside the density range proposed in literature.}
	\label{tab:tab1}
	\resizebox{1.7\columnwidth}{!}{
	\begin{tabular}{ccccccccc} 
		\hline
	Model & $\%$ice  & $\%$clathrate & $\%$salt & $\%$rock & $\rho$ [kg m$^{-3}$] & $K$ [W m$^{-1}$ K$^{-1}$] & $\mu$ [Pa s] & Ra\\
	M1 & 30 & 40 & 20 & 10 & 1400 & 1.16 & 10$^{14}$ & $10^{5}$\\
	M2 & 40 & 50 & 5  & 5  & 1150 & 1.27 & 10$^{13}$-10$^{15}$ & $10^{4}$-$10^{8}$\\
		\hline
	\end{tabular}}
\end{table*}

\subsection{Model M1}
\begin{figure*}
\includegraphics[width=1\textwidth]{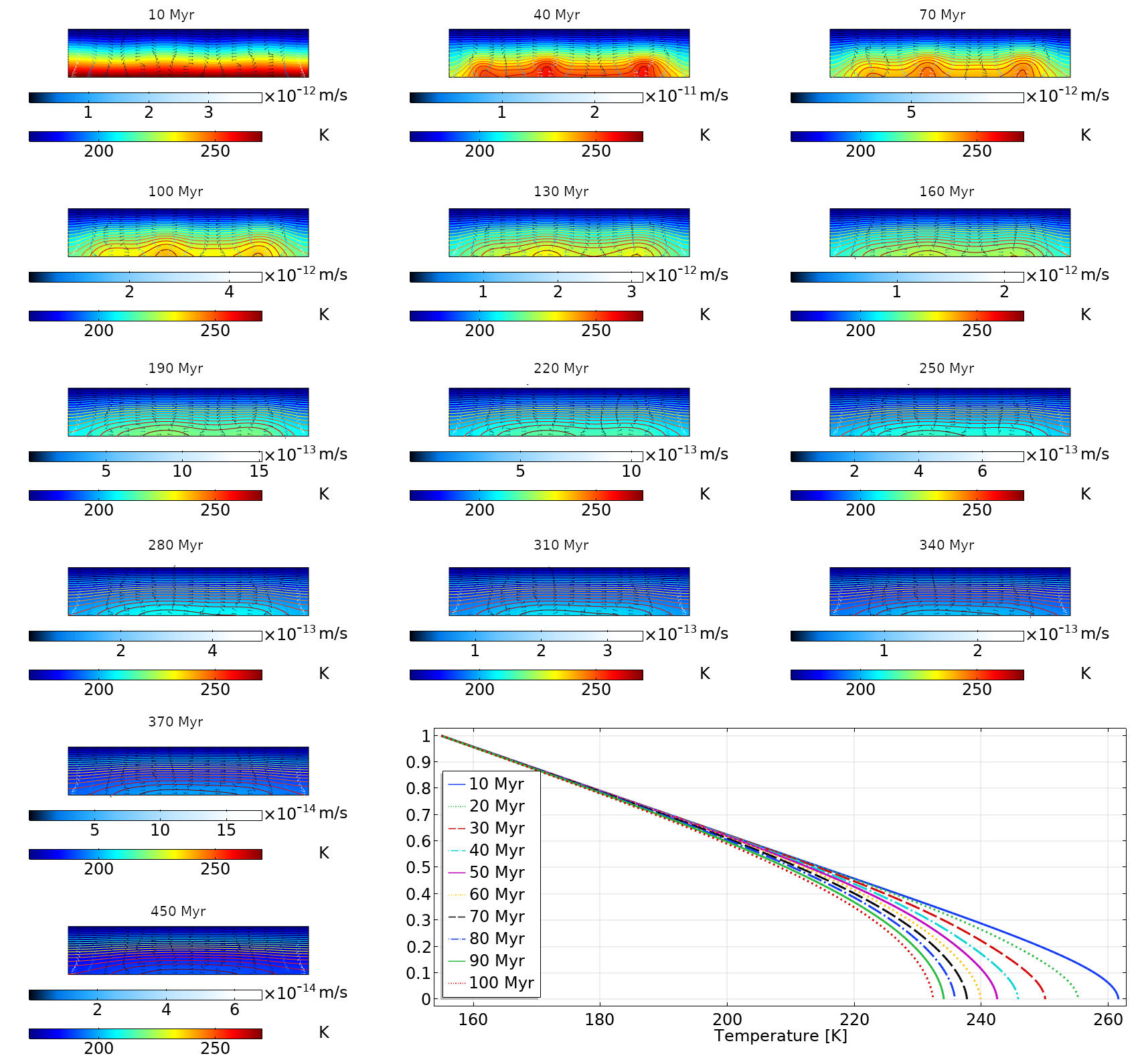}
\caption{Case M1: temperature maps from 10 Myr to 450 Myr at different times. Top colour scale refers to the convective velocity (m s$^{-1}$), while bottom colour scale refers to the temperature (K). Last panel (bottom right) shows the 1-D temperature profile at the center of the cell, at given times: it is a vertical profile, where 0 represents the bottom of the crust while 1 the surface. In the x-axis we report the temperature (K).}
     \label{fig:F1}
 \end{figure*}

We start the discussion with the model  M1, similar to the one discussed in \citet{Castillo19}: 30 vol.$\%$ ice, 40 vol.$\%$ clathrate, 20 vol.$\%$ salt and 10 vol.$\%$ rock. This composition is characterized by a mean density of 1400 kg m$^{-3}$, exactly in the middle of the range proposed by \citet{Konopliv18}, while the mean thermal conductivity is 1.16 W m$^{-1}$ K$^{-1}$. 
In Fig.\ref{fig:F1} we show the temperature maps from 10 Myr to 450 Myr after convection onset, while the last panel (bottom right) represents the 1-D temperature plot at the center of the convective cell at given times. As shown in the first three panels of Fig.\ref{fig:F1} thermal convection lasts for less 100 Myr, with a maximum convective velocity of the order of 10$^{-11}$ m s$^{-1}$. Thermal convection is very weak as also shown by the 1-D temperature plot in which it is not recognizable a clear convective isothermal profile (along the y-axis 0 represents the interface mantle-crust while 1 the top of the crust).

The initial Rayleigh number is of the order of 10$^5$ (see Fig.\ref{fig:FA2} in the Appendix) while the initial dynamic viscosity is of the order of 10$^{14}$ Pa s. 
We recall that small values of the viscosity imply high values of the Rayleigh number and consequently a vigorous thermal convection. In fact, the definition of the Rayleigh number is $g \alpha L^3 \Delta T / \kappa \nu$ where $g$ is the gravitational acceleration, $\alpha$ the thermal expansion, $L$ the size of the shell, $\Delta T$ the temperature difference across the shell, $\kappa$ the thermal diffusivity and finally $\nu$ is the kinematic viscosity. If the Rayleigh number exceeds a critical value (which depends on the physical situation under investigation, see e.g. \citet{Schubert,Solomatov07}) the initial temperature perturbation will grow with time and thermal convection starts, otherwise disturbances will decay with time. In the Appendix we report the Rayleigh number vs time plot in comparison with the critical value.
From Fig.\ref{fig:F1}, it is clear that thermal convection affects the 50$\%$ of the cell, inducing a negligible thermal stress on the surface.

\begin{figure*}
\includegraphics[width=1\textwidth]{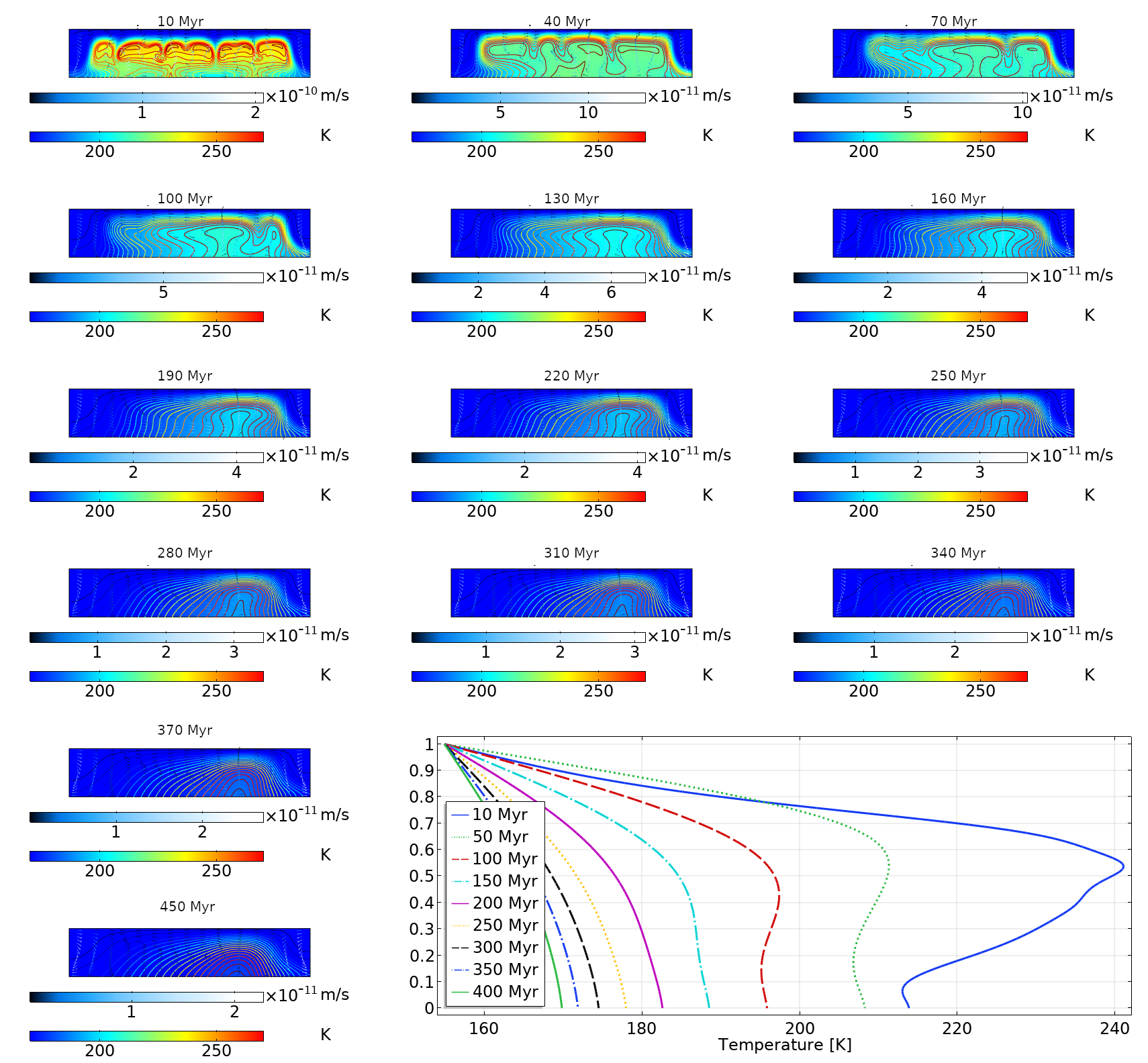}
\caption{Case M2A ($\mu_{ice} = 10^{12}$ Pa s): temperature maps from 10 Myr to 450 Myr at different times. Top colour scale refers to the convective velocity (m s$^{-1}$), while bottom colour scale refers to the temperature (K). Last panel (bottom right) shows the 1-D temperature profile at the center of the cell, at given times: it is a vertical profile, where 0 represents the bottom of the crust while 1 the surface. In the x-axis we report the temperature (K).}
     \label{fig:F2}
 \end{figure*}

\subsection{Model M2}
We analyze now a composition (M2) dominated by clathrate (50 vol.$\%$) and ice (40 vol.$\%$), with almost negligible salt (5 vol.$\%$) and rock (5 vol.$\%$) components. The corresponding mean density is 1150 kg m$^{-3}$, slightly below the minimum value of the range proposed in the literature. The mean thermal conductivity is now 1.27 W m$^{-1}$ K$^{-1}$. 
As mentioned before, we consider three different reference viscosity for the ice: 10$^{12}$ Pa s (case M2A), 10$^{13}$ Pa s (case M2B) and 10$^{14}$ Pa s (case M2C).

\subsubsection{M2A}
In this case we set the reference viscosity for the ice at 10$^{12}$ Pa s. The dynamic viscosity of the mixture at the beginning of the simulations is $\simeq10^{13}$ Pa s, which reflects in a Rayleigh number of $10^7$. In this case the thermal convection is quite strong and a nearly typical isothermal profile is evident after 100 Myr, involving about the 50$\%$ of the crust. Thermal convection is strongly reduced after 200 Myr.

\subsection{M2B}
This case is characterized by a reference viscosity of 10$^{13}$ Pa s. In the first stages, the dynamic viscosity is of the order of 10$^{14}$ Pa s and consequently the Rayleigh number is of the order of 10$^{6}$. Thermal convection is not vigorous and affects only less than 50 $\%$ of the crust. Small plumes rise in the first phases ($<$ 100 Myr) and after a rapid cooling start. In the 1-D temperature plot we do not recognize a clear isothermal profile, typical of the thermal convection.

\subsubsection{M2C}
The last is that with the reference viscosity equal to $10^{14}$ Pa s. The initial dynamic viscosity is 10$^{15}$ Pa s so the Rayleigh number is of the order of 10$^4$. In this case the thermal profile is slightly perturbed and a typical isothermal profile is not recognizable. We can only see small plumes in the first 100 Myr, which affects only the 20-30 $\%$ of the crust. 

\begin{figure*}
\includegraphics[width=1\textwidth]{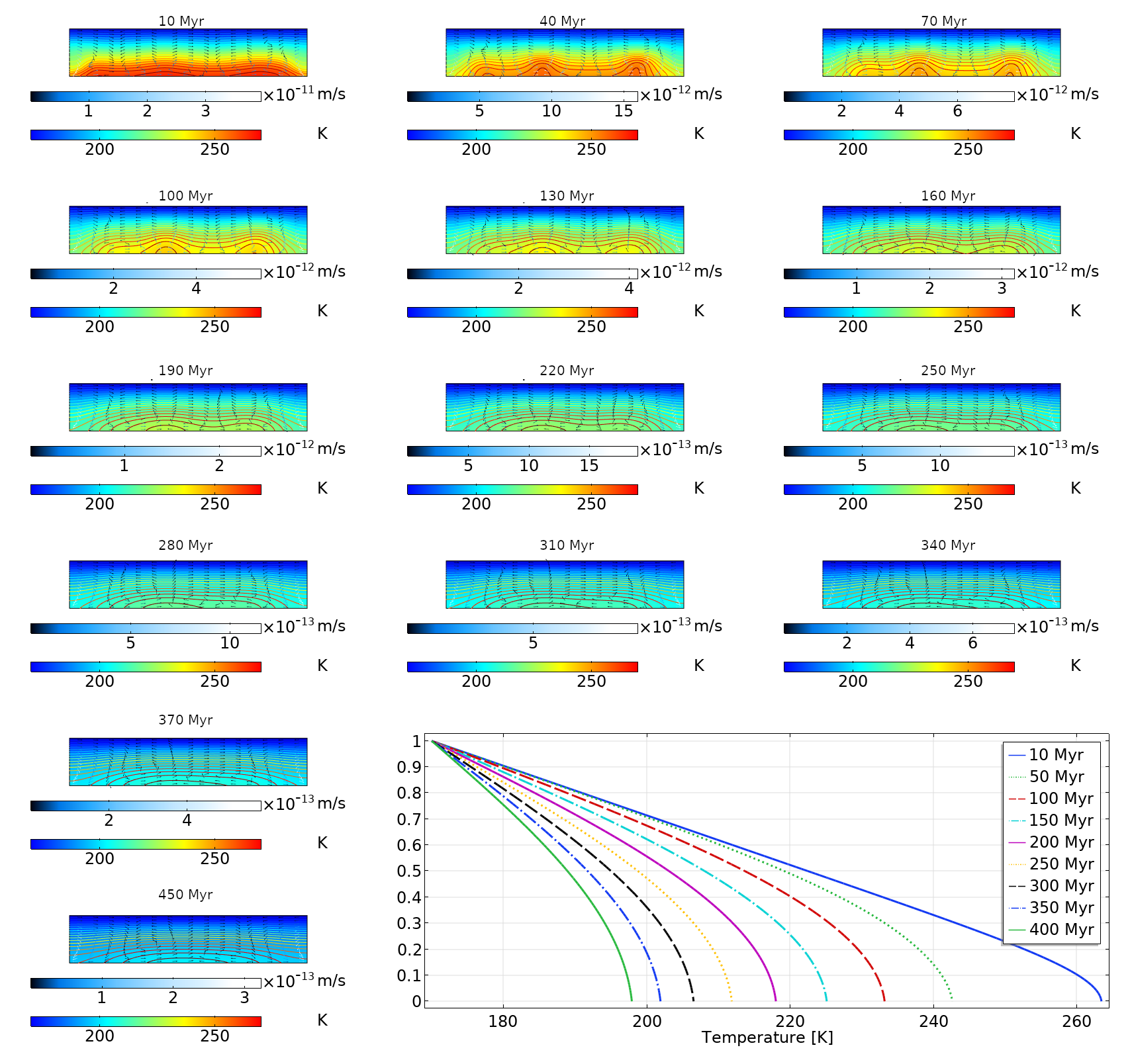}
\caption{Case M2B ($\mu_{ice} = 10^{13}$ Pa s): temperature maps from 10 Myr to 450 Myr at different times. Top colour scale refers to the convective velocity (m s$^{-1}$), while bottom colour scale refers to the temperature (K). Last panel (bottom right) shows the 1-D temperature profile at the center of the cell, at given times: it is a vertical profile, where 0 represents the bottom of the crust while 1 the surface. In the x-axis we report the temperature (K).}
     \label{fig:F3}
 \end{figure*}
 
 \begin{figure*}
\includegraphics[width=1\textwidth]{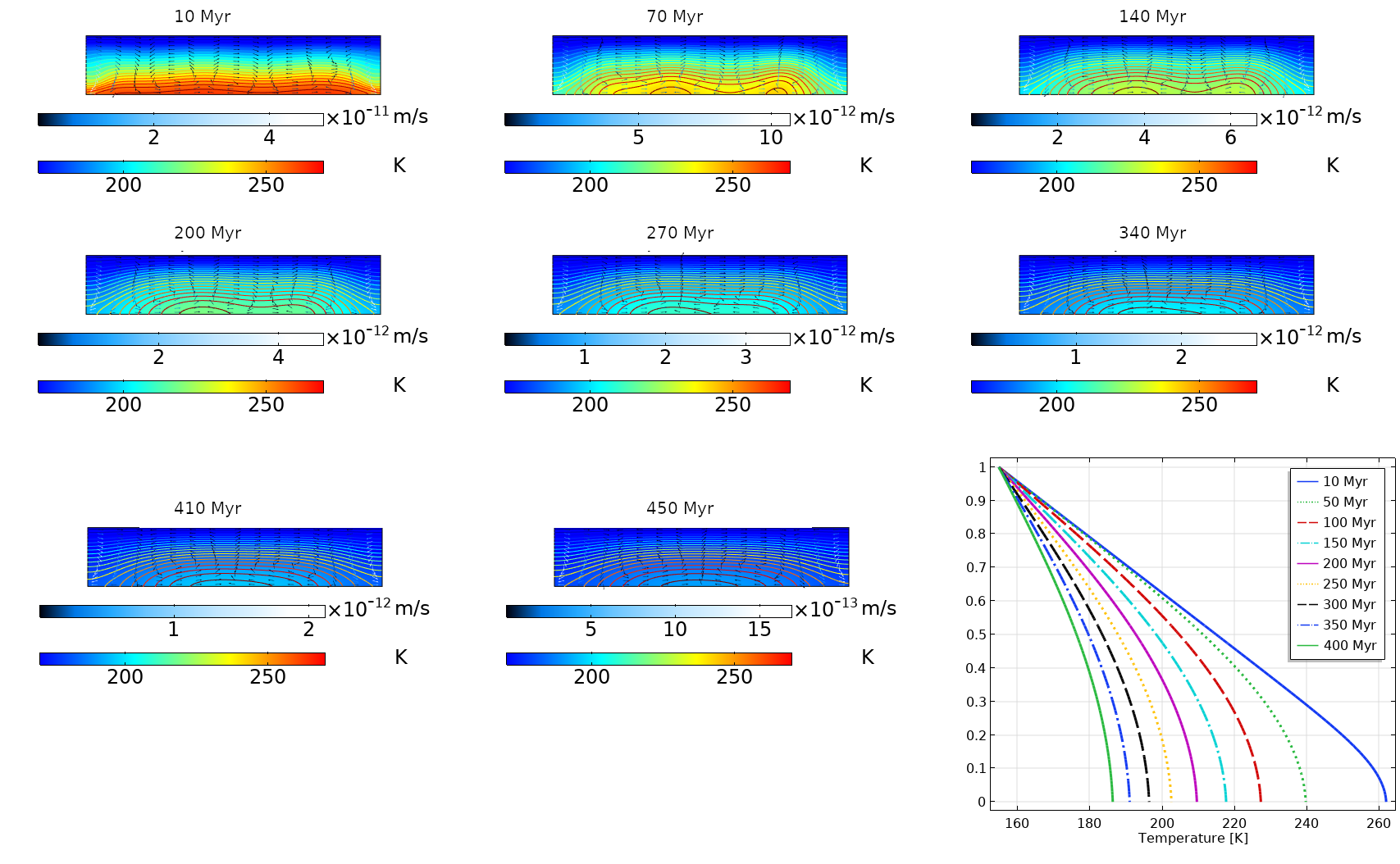}
\caption{Case M2C ($\mu_{ice} = 10^{14}$ Pa s): temperature maps from 10 Myr to 450 Myr at different times. Top colour scale refers to the convective velocity (m s$^{-1}$), while bottom colour scale refers to the temperature (K). Last panel (bottom right) shows the 1-D temperature profile at the center of the cell, at given times: it is a vertical profile, where 0 represents the bottom of the crust while 1 the surface. In the x-axis we report the temperature (K).}
     \label{fig:F4}
 \end{figure*}

\section{Discussion and Conclusions}

The numerical simulations presented in this work demonstrate that thermal convection is possible for a narrow set of thermophysical parameters. In particular, the onset of thermal convection requires a high temperature ($>$ 250 K) at the base of the crust. As pointed out by \citet{Castillo19}, such temperatures would require a crustal abundance of clathrate hydrates $>$ 55 vol.$\%$, consistent with the geological constraints of \citet{Fu2017} and \citet{Bland2016} but hard to reconcile with Ceres' impacting history since clathrate hydrates, formed during the freezing of the ocean, destabilize upon impacting. In a crust with a composition similar to those explored in \citet{Castillo19} (30 vol.$\%$ ice, 40 vol.$\%$ clathrate, 20 vol.$\%$ salt and 10 vol.$\%$ rock) thermal convection is not very vigorous, involving only the 50$\%$ of the crust. Reducing the volumetric percentage of the clathrate and  increasing that of the rock, convection is not possible.
Departing from the literature physical constraints on the crust composition, and considering clathrate (50 vol.$\%$) and ice (40 vol.$\%$) the most predominant components (salt and rock the remaining 10 vol.$\%$), thermal convection can take place.

In case of ice grain size in the range 0.1 - 0.1 mm, the thermal convection is very vigorous (Rayleigh number up to 10$^7$) involving about the 50$\%$ of the crust. In case of grain size of 1 mm, small plumes arises at the basis of the crust so the thermal convection is very weak. However, this composition (M2) has a density less than the mean crustal density obtained by e.g. \citet{Ermakov17,King18,Konopliv18}. 

Another supporting argument that  thermal convection is unlikely in Ceres' crust is provided by the experiments of \citet{Qi2018}. These authors demonstrated that just a few percent of strong particles are able to impede grain boundary sliding and we expect at least 30 vol.$\%$ of strong particles (salt and silicates) in Ceres' crust. 
It is interesting to calculate the thermal stress induced by thermal convection in order to evaluate if some of the domes observed on Ceres' surface are the product of convection. By applying a simple formula to quantify the maximum convective stress \cite{Pappalardo04}:
\begin{equation}
p \simeq 0.1 \rho g \alpha \Delta T D,
\end{equation}
where $\rho$ is the density of the outer shell, $D$ its thickness, $g$ the gravity, $\alpha$ the coefficient of thermal expansion and $\Delta T$ the gradient across the shell, we calculate that the maximum uplift of the crust, in the most favorable case, is less than 50 m. We would like to note that if the crust were 100-km thick (as suggested by past theoretical models) the maximum uplift would be $<$ 100 m \citep{FormisanoCOSPAR}. However, this estimation (even small) can be considered an upper limit. In fact \citet{Solomatov07} replaced D with $\delta$ (the thickness of the boundary layer) and treated the crust as elastic, reducing in this way the amplitude of surface topography.
This suggests that thermal convection can not produce sufficient driving pressures to produce domes like Ahuna Mons (4-km high) \citep{Ruesch16}. On the other hand, diapirism \cite{Nimmo02,Pappalardo04} could produce structure like Ahuna Mons. The possibility that in Ceres a compositional density difference produced a diapir was discussed for the first time by \citet{MCCord05}. They proposed that deep materials could reach the surface if the crust was weakened by impacts or faulting. \citet{Shoji} predicted overturn by diapirism over Ceres' history but only near its equator while \citet{Neveu15}  suggested a complete overturn of the crust linked to compositional diapirism from a subsurface ocean in case Ceres formed in 4-7 Myr from the Calcium - Alluminum - Inculsions (CAIs) formation.
The fact that thermal convection is difficult to establish in Ceres' crust could support the idea of preservation of liquid water until present, as suggested by \citet{Castillo19} and recently supported also by a numerical model \citep{Neumann20}.
However, if locally the composition was dominated by ice and clathrate, as also recently proposed by the work of \citet{Sori2020}, thermal convection would be possible.

For the presence of water ice and/or liquid Ceres represents an interesting astrobiological target to be explored by future space missions. 

\bibliography{BIBLIO}

\begin{thebibliography}{}
\makeatletter
\relax
\def\mn@urlcharsother{\let\do\@makeother \do\$\do\&\do\#\do\^\do\_\do\%\do\~}
\def\mn@doi{\begingroup\mn@urlcharsother \@ifnextchar [ {\mn@doi@}
  {\mn@doi@[]}}
\def\mn@doi@[#1]#2{\def\@tempa{#1}\ifx\@tempa\@empty \href
  {http://dx.doi.org/#2} {doi:#2}\else \href {http://dx.doi.org/#2} {#1}\fi
  \endgroup}
\def\mn@eprint#1#2{\mn@eprint@#1:#2::\@nil}
\def\mn@eprint@arXiv#1{\href {http://arxiv.org/abs/#1} {{\tt arXiv:#1}}}
\def\mn@eprint@dblp#1{\href {http://dblp.uni-trier.de/rec/bibtex/#1.xml}
  {dblp:#1}}
\def\mn@eprint@#1:#2:#3:#4\@nil{\def\@tempa {#1}\def\@tempb {#2}\def\@tempc
  {#3}\ifx \@tempc \@empty \let \@tempc \@tempb \let \@tempb \@tempa \fi \ifx
  \@tempb \@empty \def\@tempb {arXiv}\fi \@ifundefined
  {mn@eprint@\@tempb}{\@tempb:\@tempc}{\expandafter \expandafter \csname
  mn@eprint@\@tempb\endcsname \expandafter{\@tempc}}}

\bibitem[\protect\citeauthoryear{{Bland} et~al.,}{{Bland}
  et~al.}{2016}]{Bland2016}
{Bland} M.~T.,  et~al., 2016, Nature Geoscience, 9, 538

\bibitem[\protect\citeauthoryear{{Burruss}}{{Burruss}}{1981}]{Burruss1981}
{Burruss} R.,  1981, Short Course Handbook 6, Mineral. Assoc. Canada.
L.Hollister, M.L. Crawford

\bibitem[\protect\citeauthoryear{Carrozzo et~al.,}{Carrozzo
  et~al.}{2018}]{Carrozzo2018}
Carrozzo F.~G.,  et~al., 2018, Science Advances, 4

\bibitem[\protect\citeauthoryear{{Castillo-Rogez} \& {McCord}}{{Castillo-Rogez}
  \& {McCord}}{2010}]{Castillo2010}
{Castillo-Rogez} J.~C.,  {McCord} T.~B.,  2010, \icarus, 205, 443

\bibitem[\protect\citeauthoryear{Castillo-Rogez, Neveu, McSween, Fu, Toplis  \&
  Prettyman}{Castillo-Rogez et~al.}{2018}]{Castillo2018}
Castillo-Rogez J.,  Neveu M.,  McSween H.~Y.,  Fu R.~R.,  Toplis M.~J.,
  Prettyman T.,  2018, Meteoritics \& Planetary Science, 53, 1820

\bibitem[\protect\citeauthoryear{{Castillo-Rogez}, Hesse, Formisano, Sizemore,
  Bland, Ermakov  \& Fu}{{Castillo-Rogez} et~al.}{2019}]{Castillo19}
{Castillo-Rogez} J.~C.,  Hesse M.~A.,  Formisano M.,  Sizemore H.,  Bland M.,
  Ermakov A.~I.,   Fu R.~R.,  2019, Geophysical Research Letters, 46, 1963

\bibitem[\protect\citeauthoryear{{Ciarniello} et~al.}{{Ciarniello}
  et~al.}{2017}]{Ciarniello17}
{Ciarniello} M.,  et~al., 2017, Astronomy $\&$ Astrophysics, 598, A130

\bibitem[\protect\citeauthoryear{Combe et~al.,}{Combe et~al.}{2016}]{Combe16}
Combe J.-P.,  et~al., 2016, Science, 353

\bibitem[\protect\citeauthoryear{Combe et~al.,}{Combe et~al.}{2019}]{Combe18}
Combe J.-P.,  et~al., 2019, Icarus, 318, 22

\bibitem[\protect\citeauthoryear{{De Sanctis} et~al.}{{De Sanctis}
  et~al.}{2016}]{DeSanctis16}
{De Sanctis} M.,  et~al., 2016, Nature, 536, 54

\bibitem[\protect\citeauthoryear{Durham, Kirby, Stern  \& Zhang}{Durham
  et~al.}{2003}]{Durham2003}
Durham W.~B.,  Kirby S.~H.,  Stern L.~A.,   Zhang W.,  2003, Journal of
  Geophysical Research: Solid Earth, 108, 2182

\bibitem[\protect\citeauthoryear{Durham, Prieto-Ballesteros, Goldsby  \&
  Kargel}{Durham et~al.}{2010}]{Durham2010}
Durham W.~B.,  Prieto-Ballesteros O.,  Goldsby D.~L.,   Kargel J.~S.,  2010,
  Space Science Reviews, 153, 273

\bibitem[\protect\citeauthoryear{{Ermakov} et~al.,}{{Ermakov}
  et~al.}{2017}]{Ermakov17}
{Ermakov} A.~I.,  et~al., 2017, Journal of Geophysical Research (Planets), 122,
  2267

\bibitem[\protect\citeauthoryear{{Formisano}, {Federico}, {De Angelis}, {De
  Sanctis}  \& {Magni}}{{Formisano} et~al.}{2016}]{Formisano16b}
{Formisano} M.,  {Federico} C.,  {De Angelis} S.,  {De Sanctis} M.,   {Magni}
  G.,  2016, MNRAS, 463, 520

\bibitem[\protect\citeauthoryear{{Formisano}, {Federico}, {Magni}  \& {De
  Sanctis}}{{Formisano} et~al.}{2018}]{FormisanoCOSPAR}
{Formisano} M.,  {Federico} C.,  {Magni} G.,   {De Sanctis} M.~C.,  2018, in
  42nd COSPAR Scientific Assembly. pp B1.1--20--18

\bibitem[\protect\citeauthoryear{Fowler, Howell  \& Khaleque}{Fowler
  et~al.}{2016}]{Fowler16}
Fowler A.~C.,  Howell P.~D.,   Khaleque T.~S.,  2016, Geophysical \&
  Astrophysical Fluid Dynamics, 110, 130

\bibitem[\protect\citeauthoryear{Freeman}{Freeman}{2006}]{Freeman2006}
Freeman J.,  2006, \mn@doi [Planetary and Space Science]
  {https://doi.org/10.1016/j.pss.2005.10.003}, 54, 2

\bibitem[\protect\citeauthoryear{Friedson \& Stevenson}{Friedson \&
  Stevenson}{1983}]{Friedson1983}
Friedson A.,  Stevenson D.,  1983, \mn@doi [Icarus]
  {https://doi.org/10.1016/0019-1035(83)90124-0}, 56, 1

\bibitem[\protect\citeauthoryear{Fu et~al.,}{Fu et~al.}{2017}]{Fu2017}
Fu R.~R.,  et~al., 2017, Earth and Planetary Science Letters, 476, 153

\bibitem[\protect\citeauthoryear{Goldsby \& Kohlstedt}{Goldsby \&
  Kohlstedt}{2001}]{Goldsby2001}
Goldsby D.,  Kohlstedt D.,  2001, \mn@doi [Journal of Geophysical Research:
  Solid Earth] {10.1029/2000JB900336}, 106, 11017

\bibitem[\protect\citeauthoryear{Grimm \& Mcsween}{Grimm \&
  Mcsween}{1989}]{Grimm1989}
Grimm R.~E.,  Mcsween H.~Y.,  1989, Icarus, 82, 244

\bibitem[\protect\citeauthoryear{Grindrod, Fortes, Nimmo, Feltham, Brodholt  \&
  Vocadlo}{Grindrod et~al.}{2008}]{Grindrod2008}
Grindrod P.,  Fortes A.,  Nimmo F.,  Feltham D.,  Brodholt J.,   Vocadlo L.,
  2008, Icarus, 197, 137

\bibitem[\protect\citeauthoryear{Khaleque, Fowler, Howell  \&
  Vynnycky}{Khaleque et~al.}{2015}]{Fowler15}
Khaleque T.,  Fowler A.,  Howell P.,   Vynnycky M.,  2015, Physics of Fluids,
  27, 076603

\bibitem[\protect\citeauthoryear{King, Castillo-Rogez, Toplis, Bland, Raymond
  \& Russell}{King et~al.}{2018}]{King18}
King S.~D.,  Castillo-Rogez J.~C.,  Toplis M.~J.,  Bland M.~T.,  Raymond C.~A.,
    Russell C.~T.,  2018, Meteoritics \& Planetary Science, 53, 1999

\bibitem[\protect\citeauthoryear{Konopliv et~al.,}{Konopliv
  et~al.}{2018}]{Konopliv18}
Konopliv A.,  et~al., 2018, Icarus, 299, 411

\bibitem[\protect\citeauthoryear{{McCord} \& {Sotin}}{{McCord} \&
  {Sotin}}{2005}]{MCCord05}
{McCord} T.~B.,  {Sotin} C.,  2005, Journal of Geophysical Research (Planets),
  110, E05009

\bibitem[\protect\citeauthoryear{{McCord}, {Castillo-Rogez}  \&
  {Rivkin}}{{McCord} et~al.}{2011}]{McCord2011}
{McCord} T.~B.,  {Castillo-Rogez} J.,   {Rivkin} A.,  2011, \ssr, 163, 63

\bibitem[\protect\citeauthoryear{McSween~Jr. et~al.,}{McSween~Jr.
  et~al.}{2018}]{McSween18}
McSween~Jr. H.~Y.,  et~al., 2018, Meteoritics \& Planetary Science, 53, 1793

\bibitem[\protect\citeauthoryear{Nagel, Breuer  \& Spohn}{Nagel
  et~al.}{2004}]{Nagel2004}
Nagel K.,  Breuer D.,   Spohn T.,  2004, \mn@doi [Icarus]
  {https://doi.org/10.1016/j.icarus.2003.12.019}, 169, 402

\bibitem[\protect\citeauthoryear{Neumann, Jaumann, Castillo-Rogez, Raymond  \&
  Russell}{Neumann et~al.}{2020}]{Neumann20}
Neumann W.,  Jaumann R.,  Castillo-Rogez J.,  Raymond C.,   Russell C.,  2020,
  \mn@doi [A\&A] {10.1051/0004-6361/201936607}, 633, A117

\bibitem[\protect\citeauthoryear{Neveu \& Desch}{Neveu \&
  Desch}{2015}]{Neveu15}
Neveu M.,  Desch S.~J.,  2015, Geophysical Research Letters, 42, 10,197

\bibitem[\protect\citeauthoryear{Nimmo \& Manga}{Nimmo \&
  Manga}{2002}]{Nimmo02}
Nimmo F.,  Manga M.,  2002, Geophysical Research Letters, 29, 24

\bibitem[\protect\citeauthoryear{Pappalardo \& Barr}{Pappalardo \&
  Barr}{2004}]{Pappalardo04}
Pappalardo R.~T.,  Barr A.~C.,  2004, Geophysical Research Letters, 31

\bibitem[\protect\citeauthoryear{{Park} et~al.,}{{Park} et~al.}{2016}]{Park16}
{Park} R.,  et~al., 2016, in Lunar and Planetary Science Conference. p.~1781

\bibitem[\protect\citeauthoryear{{Prettyman} et~al.,}{{Prettyman}
  et~al.}{2017}]{Prettyman2017}
{Prettyman} T.~H.,  et~al., 2017, Science, 355, 55

\bibitem[\protect\citeauthoryear{Qi, Stern, Pathare, Durham  \& Goldsby}{Qi
  et~al.}{2018}]{Qi2018}
Qi C.,  Stern L.~A.,  Pathare A.,  Durham W.~B.,   Goldsby D.~L.,  2018,
  Geophysical Research Letters, 45, 12,757

\bibitem[\protect\citeauthoryear{Raponi et~al.,}{Raponi
  et~al.}{2018}]{Raponi2018a}
Raponi A.,  et~al., 2018, Science Advances, 4

\bibitem[\protect\citeauthoryear{Rubin, Desch  \& Neveu}{Rubin
  et~al.}{2014}]{Rubin2014}
Rubin M.~E.,  Desch S.~J.,   Neveu M.,  2014, \mn@doi [Icarus]
  {https://doi.org/10.1016/j.icarus.2014.03.047}, 236, 122

\bibitem[\protect\citeauthoryear{Rudman}{Rudman}{1992}]{Rudman1992}
Rudman M.,  1992, \mn@doi [Physics of the Earth and Planetary Interiors]
  {https://doi.org/10.1016/0031-9201(92)90199-6}, 72, 153

\bibitem[\protect\citeauthoryear{Ruesch et~al.,}{Ruesch
  et~al.}{2016}]{Ruesch16}
Ruesch O.,  et~al., 2016, Science, 353

\bibitem[\protect\citeauthoryear{{Schmidt} et~al.}{{Schmidt}
  et~al.}{2017}]{Schmidt}
{Schmidt} B.~E.,  et~al., 2017, Nature Geoscience, 10, 338

\bibitem[\protect\citeauthoryear{{Schofield}, {Alsop}, {Warren}, {Underhill},
  {Lehne}, {Beer}  \& {Lukas}}{{Schofield} et~al.}{2014}]{Schofield2014}
{Schofield} N.,  {Alsop} I.,  {Warren} J.,  {Underhill} J.~R.,  {Lehne} R.,
  {Beer} W.,   {Lukas} V.,  2014, Geology, 42, 599

\bibitem[\protect\citeauthoryear{Shoji \& Kurita}{Shoji \&
  Kurita}{2014}]{Shoji}
Shoji D.,  Kurita K.,  2014, Journal of Geophysical Research: Planets, 119,
  2457

\bibitem[\protect\citeauthoryear{Solomatov \& Barr}{Solomatov \&
  Barr}{2006}]{Solomatov07}
Solomatov V.,  Barr A.,  2006, Physics of the Earth and Planetary Interiors,
  155, 140

\bibitem[\protect\citeauthoryear{Solomatov \& Moresi}{Solomatov \&
  Moresi}{2000}]{Solomatov2000}
Solomatov V.~S.,  Moresi L.-N.,  2000, \mn@doi [Journal of Geophysical
  Research: Solid Earth] {10.1029/2000JB900197}, 105, 21795

\bibitem[\protect\citeauthoryear{{Sori}, {Sizemore}, {Byrne}, {Bramson},
  {Bland}, {Stein}  \& {Russell}}{{Sori} et~al.}{2018}]{Sori18}
{Sori} M.~M.,  {Sizemore} H.~G.,  {Byrne} S.,  {Bramson} A.~M.,  {Bland} M.~T.,
   {Stein} N.~T.,   {Russell} C.~T.,  2018, Nature Astronomy, 2, 946

\bibitem[\protect\citeauthoryear{{Sori} et~al.,}{{Sori}
  et~al.}{2020}]{Sori2020}
{Sori} M.~M.,  et~al., 2020, in Lunar and Planetary Science Conference. Lunar
  and Planetary Science Conference.
p.~1651

\bibitem[\protect\citeauthoryear{Sterenborg \& Crowley}{Sterenborg \&
  Crowley}{2013}]{Sterenborg13}
Sterenborg M.~G.,  Crowley J.~W.,  2013, \mn@doi [Physics of the Earth and
  Planetary Interiors] {https://doi.org/10.1016/j.pepi.2012.10.006}, 214, 53

\bibitem[\protect\citeauthoryear{Stevenson, Spohn  \& Schubert}{Stevenson
  et~al.}{1983}]{Stevenson83}
Stevenson D.~J.,  Spohn T.,   Schubert G.,  1983, Icarus, 54, 466

\bibitem[\protect\citeauthoryear{{Thomas}, {Reynolds}, {Squyres}  \&
  {Cassen}}{{Thomas} et~al.}{1987}]{Thomas1987}
{Thomas} P.~J.,  {Reynolds} R.~T.,  {Squyres} S.~W.,   {Cassen} P.~M.,  1987,
  in Lunar and Planetary Science Conference. Lunar and Planetary Science
  Conference.
p.~1016

\bibitem[\protect\citeauthoryear{Torquato, Truskett  \& Debenedetti}{Torquato
  et~al.}{2000}]{Torquato2000}
Torquato S.,  Truskett T.~M.,   Debenedetti P.~G.,  2000, \mn@doi [Phys. Rev.
  Lett.] {10.1103/PhysRevLett.84.2064}, 84, 2064

\bibitem[\protect\citeauthoryear{{Turcotte} \& {Schubert}}{{Turcotte} \&
  {Schubert}}{2002}]{Schubert}
{Turcotte} D.~L.,  {Schubert} G.,  2002, {Geodynamics - 2nd Edition}.
Cambridge University

\bibitem[\protect\citeauthoryear{{Vangelov} \& {Jarvis}}{{Vangelov} \&
  {Jarvis}}{1994}]{Vangelov94}
{Vangelov} V.~I.,  {Jarvis} G.~T.,  1994, Journal of Geophysical Research
  (Solid Earth), 99, 9345

\bibitem[\protect\citeauthoryear{Waite, Stern, Kirby, Winters  \& Mason}{Waite
  et~al.}{2007}]{Waite2007}
Waite W.~F.,  Stern L.~A.,  Kirby S.~H.,  Winters W.~J.,   Mason D.~H.,  2007,
  Geophysical Journal International, 169, 767

\bibitem[\protect\citeauthoryear{van Keken, Spiers, van~den Berg  \&
  Muyzert}{van Keken et~al.}{1993}]{VanKeken93}
van Keken P.,  Spiers C.,  van~den Berg A.,   Muyzert E.,  1993, \mn@doi
  [Tectonophysics] {https://doi.org/10.1016/0040-1951(93)90310-G}, 225, 457

\makeatother
\end{thebibliography}
\bibliographystyle{mnras}

\section*{Acknowledgments}
We would like to thank an anonymous referee for this suggestions and comments. 
This work is supported by an ASI (Agenzia Spaziale Italiana) grant.
Data table and video of the whole thermal history of the several models we developed are reported at https://github.com/MiFormisano/miformisano. 


\appendix
\section{Dimensionalization procedure}
\subsection{Reducing Navier-Stokes equations to non-dimensional forms}
We start from the Navier-Stokes equations.
We divide by $\rho_0$ and use a dimensionless version of the viscosity as in \citet{Fowler16}:
\begin{multline}
 \frac{\partial u}{\partial t} + \left(u \cdot \nabla u\right)  = \frac{1}{\rho_0}\nabla \cdot \left[\eta(T) \nabla u + \eta(T)\left(\nabla u\right)^T\right]-
 \frac{\nabla \cdotp pI}{\rho_0}\\ -g\left[1-\alpha\left(T-T_s\right)\right].
\end{multline}
Note that $\left(\nabla u\right)^T$ term does not contribute to the momentum equations because $\nabla \cdot \left(\nabla u\right)^T = 0$ for incompressible fluid, so:
\begin{equation}
 \frac{\partial u}{\partial t} + \left(u \cdot \nabla u\right)  = \frac{\nabla\eta(T)\nabla u}{\rho_0} + \frac{\eta(T)\nabla^2 u}{\rho_0} -
 \frac{\nabla \cdotp pI}{\rho_0} -g\left[1-\alpha\left(T-T_s\right)\right],
\end{equation}
where $\eta(T)=\eta_0 exp\left(\frac{1-T}{\varepsilon T}\right)$ and $\varepsilon=RT_b/E$. 
Substituting the expression for $\eta(T)$, introducing $\nu_0 = \eta_0 / \rho_0$, and calling for simplicity $V(T) = exp\left(\frac{1-T}{\varepsilon T}\right)$, we can rewrite the above equation as :
\begin{equation}
\frac{\partial u}{\partial t} + \left(u \cdot \nabla u\right)  = \nu_0\nabla V(T) \nabla u + \nu_0 V(T) \nabla^2 u -  \frac{\nabla \cdotp pI}{\rho_0} -g\left[1-\alpha\left(T-T_s\right)\right]
\end{equation}

We use the following non-dimensional factors, as in \citet{Fowler15,Fowler16}:
\begin{equation}
 u = \frac{k}{L}u^*;\quad T = T_bT^*; \quad \nabla = \frac{1}{L} \nabla^*; \quad t = \frac{L^2}{k}t^*.
\end{equation}

We can rewrite the Navier-Stokes equation as follows:
\begin{multline}
\frac{k^2}{L^3}\left(\frac{\partial u^*}{\partial t^*} + u^* \cdot \nabla ^* u^* \right) + \frac{ \nabla p}{\rho_0} = \frac{\nu_0 k}{L^3}\left[\nabla^*V(T)\nabla^*u^* + V(T)\nabla^{2*}u^*\right]\\
 -g +  g\alpha\left(T_bT^*-T_b\right],
\end{multline}
following \citet{Fowler15,Fowler16}, we define the pressure as:
\begin{equation}
 p = \rho_0 g L \left(1-z^*\right) + \frac{\eta_0 k}{L^2} p^*.
\end{equation}
and multiply all the terms for $\frac{L^3}{k \nu_0}$, we obtain: 
\begin{multline}
\frac{k}{\nu_0}\left(\frac{\partial u^*}{\partial t^*} + u^* \cdot \nabla^* u^* \right) + \nabla^* p^* = \left[\nabla^*V(T)\nabla^*u^* + V(T)\nabla^{2*}u^*\right] + \\\frac{g \alpha L^3}{k \nu_0} T_b\left(T^*-1\right).
\end{multline}
If we introduce the Prandtl and Rayleigh numbers, we finally obtain:
\begin{equation}
\frac{1}{Pr}\left(\frac{\partial u^*}{\partial t^*} + u^* \cdot \nabla^* u^* \right) + \nabla^* p^* = \left[\nabla^*V(T)\nabla^*u^* + V(T)\nabla^{2*}u^*\right] - Ra\left(1-T^*\right).
\end{equation}

\subsection{Reducing heat equation to non-dimensional}
We divide by $\rho$c$_p$ the heat transfer equation:
\begin{equation}
\frac{\partial T}{\partial t} + u \cdot \nabla T + \kappa \nabla^2 T = 0,  
\end{equation}
where $\kappa = \frac{K}{\rho c_p}$ is the thermal diffusivity. We use the following non-dimensional parameters:
\begin{equation}
\nabla = \frac{1}{L}\nabla^*;\quad T = T_bT^*; \quad t = \frac{L^2}{\kappa}t^*; \quad v = \frac{\kappa}{L}v^*.
\end{equation}
We can rewrite the heat equation in the following way:
\begin{equation}
\frac{\kappa T_b}{L^2}\frac{\partial T^*}{\partial t^*} + \frac{\kappa v^* T_b}{L^2}\nabla^* T^* + \frac{\kappa T_b}{L^2}\nabla^{2^*}T^* = 0.
\end{equation}
Simplifying the common terms we obtain the dimensionless form of the heat equation:
\begin{equation}
\frac{\partial T^*}{\partial t^*} + v^*\nabla^*T^* + \nabla^2{^*}T^* = 0.
\end{equation}

\section{Numerical Grid}
\begin{figure}
\includegraphics[width=\columnwidth]{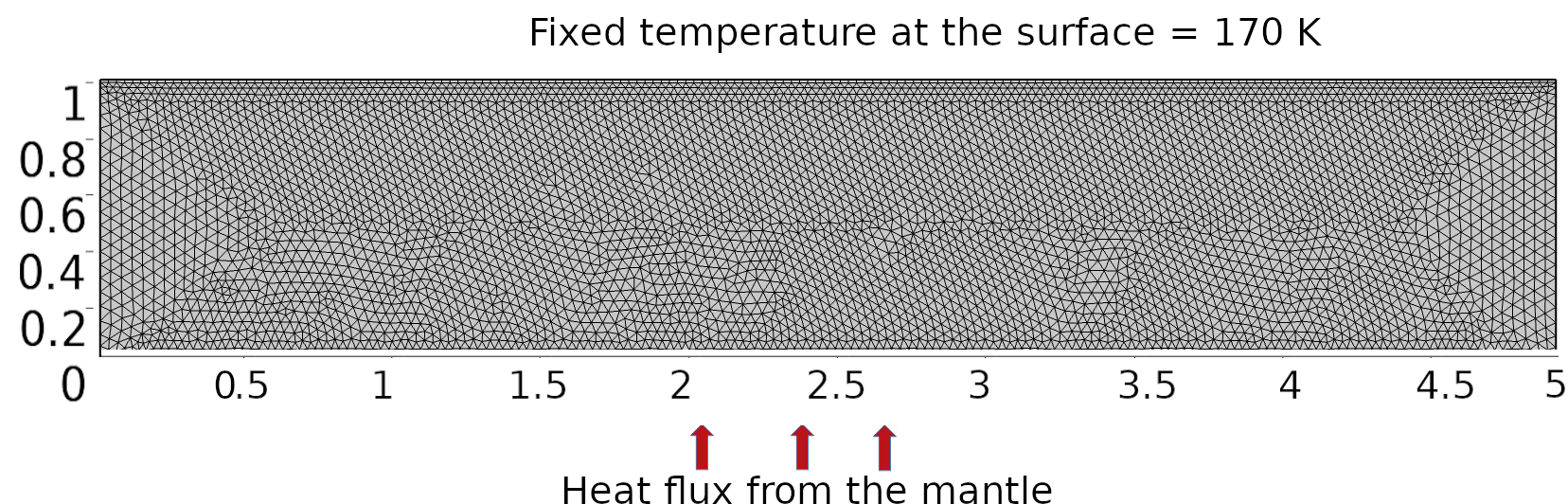}
\caption{Geometry adopted for our numerical simulations. Fixed temperature (170 K) is imposed on the top while thermal insulation on the lateral sides. An heat flux is imposed in the interface mantle-crust. Free slip conditions are imposed on the lateral sides.}
     \label{fig:FA1}
 \end{figure}

The mesh is automatically chosen by the software COMSOL Multiphysics, based on the physical equations involved. It consists (see Fig.\ref{fig:FA1}) in 10120 elements (9762 triangles and 358 quads). 
\section{Material Properties}

In Tab.\ref{tab:tab2} we report the thermophysical material properties of the materials we have used for our numerical simulations.
\begin{table}
	\centering
	\caption{Thermophysical material properties assumed in this study, in particular the density ($\rho$), the thermal conductivity ($K$), the specific heat ($c_p$) and the thermal expansion coefficient ($\alpha$). References: [1] \citet{Grindrod2008}; [2] \citet{Grimm1989}; [3] \citet{Waite2007}; [4] \citet{Schofield2014}; [5] \citet{Shoji}. * refers to the mean value over the range 170-275 K calculated using the temperature-dependent relations given in the specific references.}
	\label{tab:tab2}
	\resizebox{1\columnwidth}{!}{
	\begin{tabular}{|c|r|r|r|r|} 
		\hline
		Material & $\rho$ [kg m$^{-3}$] & $K$ [W m$^{-1}$ K$^{-1}] $ & $c_p$ [J kg$^{-1}$ K$^{-1}$] & $\alpha$ [K$^{-1}$]\\
		\hline
		Ice & 950 [1] & 2.7 [2]* & 1750 [2]* & 10$^{-4}$ [5]*\\
		\hline
		Clathrate hydrate & 1000 [3] & 0.64 [3] & 1850 [3]* & 10$^{-7}$ [1]*\\
		\hline
		Antigorite ("rock") & 2750 [1] & 1.5 [1] & 2000 [1] & 10$^{-6}$ [1]*\\ 
		\hline
		Salt (hydrohalite) & 2200 [4] & 0.60 [4] & 920 [4] & -- --\\
       	\hline
	\end{tabular}}
\end{table}

\section{Rayleigh number vs time evolution}
\begin{figure}
\includegraphics[width=\columnwidth]{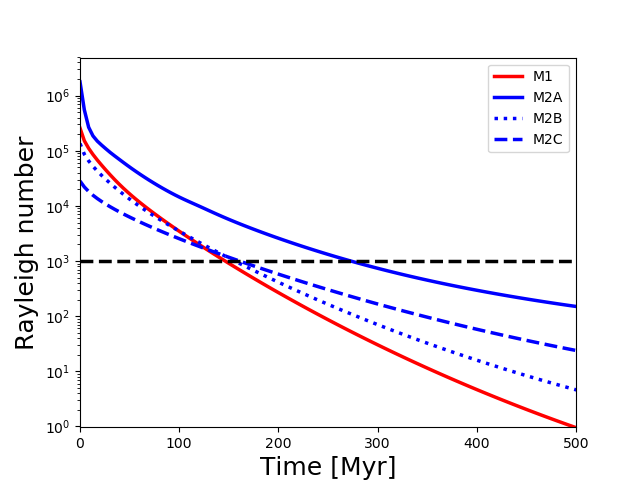}
\caption{Rayleigh number evolution for the models under investigation. Dashed horizontal black line represent the critical Rayleigh number, evaluated as in \citet{Solomatov07}.}
     \label{fig:FA2}
 \end{figure}

Here we report the plot of the Rayleigh number evolution with time. The critical Rayleigh number is calculated according to \citet{Solomatov07} and marks the onset of the thermal convection. It depends on the geometry and physics of the problem under investigation. An appropriate definition, however, is reported in \citet{Stevenson83}: "\emph{...should more correctly be thought of as an empirical parameter chosen to be consistent with numerical and laboratory experiments".} In Fig.\ref{fig:FA2} we report the Rayleigh number evolution with time for the models under investigation. The intersection with the critical value (black horizontal dashed line) gives an estimation of the timescale of the thermal convection.

\bsp	
\label{lastpage}
\end{document}